\documentclass[prl,twocolumn,aps,showpacs]{revtex4}

\usepackage{graphicx}
\usepackage{bm}
\usepackage{amsmath}

\newcommand{\ket}[1]{\left| #1 \right>} 

\begin{document}

\title{Exact Parent Lattice Hamiltonian for One-Dimensional \\ Non-Abelian Fractional Quantum Hall Liquids}

\author{Bel\'{e}n Paredes}

\affiliation{Department f\"ur Physik and Arnold Sommerfeld Center for Theoretical Physics, Ludwig-Maximilians-Universit\"at, Theresienstrasse 37 80333 M\"unchen, Germany\\}

\date{\today}

\begin{abstract} I present a family of one-dimensional time reversal lattice Hamiltonians whose exact ground states are bosonic non-Abelian fractional quantum Hall liquids with well defined total chiral momentum. These Hamiltonians describe $k$-hard-core bosons and involve long-range tunneling and interaction processes with an amplitude decaying with the chord distance. For $k=2$ the ground state is a $p$-wave paired phase corresponding to the Pfaffian quantum Hall state. This family of non-Abelian liquids is shown to be in one-to-one correspondence with a family of non-Abelian spin-$\frac{k}{2}$ liquids exhibiting hidden non-local order. They are total singlets made out of $k$ indistinguishable Resonating Valence Bond states. The corresponding spin parent Hamiltonians are obtained.
\end{abstract}

\maketitle

Non-Abelian states of matter \cite{Stern2011} represent an exotic organizational form of quantum matter that contradicts the traditional paradigms of condensed matter physics. In these states, particles are ordered following a hidden global pattern which is not associated to the breaking of any symmetry and is revealed in the presence of topological quasiparticles obeying non-Abelian statistics \cite{Stern2008,Nayak2008}. 
The study of non-Abelian phases is at the frontier of current theoretical and experimental research \cite{Stern2011} with three fundamental objectives: the understanding of the subtle interplay between topology and quantum mechanics leading to the formation on non-Abelian phases, the experimental realization of non-Abelian anyons in the laboratory, and the possible realization of a topological quantum computer \cite{Nayak2008,DasSarma2005}, in which non-Abelian anyons could be used to encode and manipulate information in a manner resistant to errors.
There is a strong belief that non-Abelian phases might exist in fractional quantum Hall systems in the form of the so called Pfaffian state \cite{MooreRead91, Greiter92, ReadRezayi99, Nayak96}, and a huge theoretical and experimental effort is currently dedicated to the observation of such state in systems at filling factor $\nu=5/2$ \cite{Dolev2008, Radu2008, DasSarma2005, SternHalperin2006, Stern2011}. Closely related states have been predicted to occur in superconductors with $p$-wave pairing symmetry \cite{Nayak2008,Ivanov2001} and hybrid systems of superconductors with so-called topological insulators \cite{Kane2008}.
It remains a challenge to develop  theoretical models that allow us to deepen our understanding of non-Abelian phases and help us to identify other experimental setups where to observe them and test the non-Abelian statistics of their excitations.

In the search for scenarios leading to the emergence of non-Abelian states, it is illuminating to analyze the essential features of non-Abelian liquids expected to appear in fractional quantum Hall systems \cite{MooreRead91, Greiter92, ReadRezayi99, Nayak96}. In these liquids, particles organize themselves forming clusters, in such a way that all clusters are described by the same many-body wave function, which corresponds to an incompressible liquid state like, for instance, the Laughlin state \cite{Laughlin83}. The indistinguishability between clusters gives rise to an internal degree of freedom in the space of excitations, which together with the incompressibility of each cluster, leads to the non-Abelian character of the quasiparticles.
The emergence of such cluster structures can be thought as the result of the interplay of three ingredients: 1) projection onto the lowest Landau level, 2) conservation of total angular momentum, and 3) commensurability, at certain special filling factors, between the number of particles and the total angular momentum that minimizes the interaction between the particles.

Projection onto the lowest Landau level \cite{MacDonald94}, where single-particles states (in the symmetric gauge) have well defined angular momentum, resembles the projection onto the lowest band of a {\em one-dimensional} lattice system \cite{Giamarchi}, where single-particle states with well defined quasimomentum have the same functional form as those of the lowest Landau level. Naively, one is tempted to predict that non-Abelian liquids equivalent to those of fractional quantum Hall systems could emerge for one-dimensional systems in the presence of dominating interactions, provided that time-reversal symmetry was broken in the appropriate manner. This possibility was conjectured in \cite{Paredes2007}. However, there is a fundamental difference between both systems which seems to endanger the stability of such one-dimensional states: for a one-dimensional lattice system, interactions do not preserve the total angular momentum in the ring, which is only conserved modulo $2\pi$. It seems that no reasonable Hamiltonian would stabilize a one-dimensional non-Abelian fractional quantum Hall liquid: such a state would rapidly decay through mixing with other states with equivalent angular momentum.

In this work I present a family of one-dimensional time reversal lattice Hamiltonians whose exact ground states are $SU(2)_k$ non-Abelian fractional quantum Hall bosonic states of the form:
\begin{equation}
\Psi^{(k)}=
\prod_{i=1}^{kn} z_i \,\,
{\mathcal{S}} \left\{
\Phi [{z^{(1)}_1,\dotsc,z^{(1)}_{n}}] \,
\dotsm \,
\Phi [z^{(k)}_1,\dotsc,z^{(k)}_{n}] \right\}.
\label{ClusterStates}
\end{equation}
These states consist of $k$ identical copies of the Laughlin state
\begin{equation}
\Phi[z]= \prod_{i<j} (z_i-z_j)^2,
\label{LaughlinState}
\end{equation}
symmetrized over all possible assignments of the $kn$ particles to each copy.  The coordinate $z=e^{i2\pi x/M}$ is the one-dimensional complex coordinate in the ring, with $x=1,\ldots, M$ denoting the lattice site and $M$ the number of lattice sites. I consider the case in which $M=2n$, so that the filling factor of the state $\Psi^{(k)}$ is $\nu=k/2$.
For $k=2$ this state corresponds to the celebrated Pfaffian state:
\begin{equation}
\Psi^{(2)}= \prod_{i} z_i \,\text{Pff}\left(\frac{1}{z_i-z_j}\right) \prod_{i<j} (z_i-z_j),
\label{PfaffianState}
\end{equation}
which is believed to underlie the physics of fractional quantum Hall systems at filling factor $5/2$. At level $k$ the state (\ref{ClusterStates}) corresponds to a non-Abelian state holding $SU(2)_k$ non-Abelian anyons \cite{ReadRezayi99,Nayak96}. The idea of using symmetrized indistinguishable cluster states as Ansatz wave-functions for non-Abelian one-dimensional bosonic liquids was proposed by the author in \cite{Paredes2007}.  
\begin{figure}[t!]
\includegraphics[width=\linewidth]{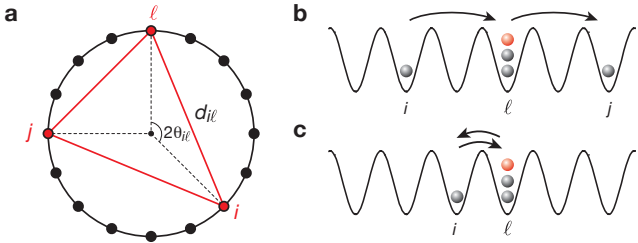}
\caption{\textbf{Three-body Hamiltonian for $k=2$ hard-core bosons.} 
\textbf{a,} One-dimensional lattice ring with $M$ sites. The chord distance between sites $i$ and $\ell$ is given by $d_{i\ell}=M/\pi\sin \theta_{i\ell}$. \textbf{b}, \textbf{c}, Schematics of virtual processes for $k=2$ hard-core bosons leading to long-range tunneling (\textbf{b}) and interactions (\textbf{c}).}
\label{SchemeHamiltonian}
\end{figure}

The one-dimensional states in Eq.\,(\ref{ClusterStates}) can not be characterized by any local order parameter. They exhibit a global hidden order which is associated to the organization of the bosonic particles in identical indistinguishable clusters. I show below that at level $k$, the state in  Eq. \,(\ref{ClusterStates}) can be uniquely characterized by being the only 
one-dimensional bosonic state satisfying the following three conditions at the same time:
1) it vanishes when $k$ particles coincide at the same lattice site, 2) it possesses time reversal symmetry, and 3) it has well defined total {\em chiral momentum}, a quantum number that is defined in this work in analogy to the angular momentum in fractional quantum Hall systems. 
By imposing conservation of these global quantum numbers an exact parent Hamiltonian emerges, which has the the form:
\begin{equation}
H^{(k)}_B=
\sum_{ \ell \neq i, \ell \neq j}
\frac{\cos \theta_{ij}}{\sin \theta_{i \ell} \sin \theta_{j \ell}} 
a^{\dagger(k)}_{i} n^{(k)}_{\ell} a^{(k)}_{j}.
\label{Ham3Body}
\end{equation}
Here the operators $a_i^{(k)}(a_i^{\dagger (k)})$ are bosonic annihilation (creation) operators at lattice site $i$, constrained to the $k$-hard core condition $[a_i^{(k)}]^{k+1}=0$. 
The operator $n^{(k)}_\ell=(1-[a_\ell^{(k)},a_\ell^{\dagger (k)}])/(k+1)$ vanishes except when the site $\ell$ is maximally occupied with $k$ particles. The Hamiltonian in Eq.\,(\ref{Ham3Body}) involves two type of processes. On the one hand, it involves long-range tunneling between sites $i$ and $j$ mediated by a third site $\ell$ that is maximally occupied with $k$ particles. The amplitude of these processes, $J_{i\ell j}=\cos \theta_{ij}/\sin \theta_{i \ell} \sin \theta_{j \ell}$, depends on the three angles of the triangle determined by the sites $i,\ell,j$. This amplitude decays as $1/d_{i\ell}d_{j\ell}$, where $d_{ij}$ is the chord distance between the sites $i$ and $j$ (see Fig.\,\ref{SchemeHamiltonian}). On the other hand, the Hamiltonian involves long range density-density interactions between two sites (terms in Eq.\,(\ref{Ham3Body}) with $i=j$), provided that one of them is maximally occupied. The interaction coupling is $K_{i\ell}=J_{i\ell i}=\sin^{-2}\theta_{i\ell}\propto d_{i\ell}^{-2}$ \cite{Note}.

\begin{figure}[t!]
\includegraphics[width=\linewidth]{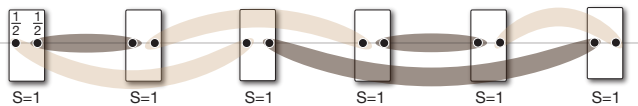}
%
\caption{\textbf{Schematics of Pfaffian Resonating Valence Bond State}. 
The spin-$1$ Pfaffian liquid can be thought as a state of fictitious on-site pairs of $1/2$ spins. These $1/2$ spins are distributed in two identical RVB states. Each local pair is then projected onto the subspace of total spin $1$.}

\label{RVB}

\end{figure}

The family of states $\Psi^{(k)}$ is shown to be in one-to-one correspondence with a family of non-Abelian spin-$\frac{k}{2}$ total singlet states of the form:
\begin{equation}
\ket{\Psi_{\text{spin}=k/2}}=\!\!\!
\sum_{x_1,\dotsc, x_{kn}} \!
\Psi^{(k)}[z_1,\dotsc,z_{kn}] \,\,\,
\widetilde{S}^+_{x_1} \dotsm \widetilde{S}^+_{x_{kn}} \ket{-},
\label{SpinState}
\end{equation}
where $\widetilde{S}^+_{x}$ is a renormalized spin-$\frac{k}{2}$ raising operator of the spin at site $x$, and
$\ket{-}=\otimes_{i=1}^M \ket{-k/2}_i$ is the state with well defined minimum total $S^z$. For $k=2$ the  spin state in Eq.\,\ref{SpinState} has been proposed as an Ansatz wave function for a non-Abelian spin liquid in two dimensions \cite{Thomale2009}.
The mapping I present here between hard-core bosons and spins allows to derive a parent Hamiltonian for these spin liquids by transforming the bosonic Hamiltonian (\ref{Ham3Body}) into a rotationally invariant spin Hamiltonian. For $k=1$ the state (\ref{SpinState}) is the Jastrow Resonating Valence Bond (RVB) state proposed by Haldane \cite{Haldane88}, a superposition of many different arrangements in which the spins are paired forming long-range singlets. The corresponding spin Hamiltonian is the Haldane-Shastry Hamiltonian \cite{Haldane88}. It is illuminating to see how Haldane-Shastry Hamiltonian emerges purely from the conservation of two global quantum numbers: the total chiral momentum and the total spin. 
For $k\ge 2$ the state (\ref{SpinState}) has hidden topological order, which is revealed by writing the state as a tensorial product of $k$ Haldane RVB states in the form:
\begin{align}
\ket{\Psi_{\text{spin}=k/2}}=\prod_{i=1}^M \mathcal{P}_i^{(\text{s}=k/2)} \ket{\Psi_{\text{s}_1=1/2}} \otimes \dotsm \otimes \ket{\Psi_{\text{s}_k=1/2}}.
\label{ClusterRVB}
\end{align}
Here $\mathcal{P}_i^{(\text{s}=k/2)}$ is a local projector that maps the tensorial product space of $k$ ficticious $1/2$ spins at site $i$ onto the completely symmetric subspace with total spin $k/2$. 
For $k=2$ this state is depicted in Fig.\,(\ref{RVB}). It resembles the AKLT spin-$1$ state \cite{Auerbach}. Here the effective $1/2$ spins are paired forming long-range singlets, giving rise to two identical RVB states. This state has hidden $Z_2\times Z_2$ symmetry,  owing to the $Z_2$ symmetry of each of the RVB states.
The corresponding parent Hamiltonian involves three-spin interactions of the form:
\begin{equation}
H_{\text{spin}=1}^{(3)}=
\sum_{i \ne j \ne \ell} \frac{\cos \theta_{ij}}{\sin \theta_{i \ell} \sin \theta_{j\ell}} 
\left(\vec{S}_i \cdot \vec{S}_\ell \right) \left(\vec{S}_\ell \cdot \vec{S}_j\right),
\label{HamSpin1}
\end{equation}
where $S^\alpha$, $\alpha=x,y,z$ are spin-$1$ operators.

For $k=2$ the properties of the Pfaffian bosonic state presented here resemble those of a Haldane insulator \cite{Altman2006, Altman2008, Endres2011}, in which string order emerges as a result of correlations between particle-hole excitations. Moreover, the spin-$1$ Hamiltonian (\ref{HamSpin1}) is similar to the Haldane biquadratic Hamiltonian for a quantum spin-$1$ chain \cite{Haldane83}.

{\em Many-body states for $k$-hard-core bosons with well defined total chiral momentum}.
Let me consider a system of bosonic particles in a one-dimensional lattice, and the basis of single-particle states with well defined quasimomentum  $\ket{q}=b^{\dagger}_q \ket{0}$, where $\ket{0}$ is the vacuum state, $b^{\dagger}_q=\frac{1}{\sqrt{M}}\sum_x z^q a^{\dagger}_x$, $z=e^{i2\pi x/M}$
is the complex coordinate in the ring corresponding to site $x=1,\dotsm, M$, and $q=0,\dotsm, M-1$.
Let me define the chiral momentum operator as:
\begin{equation}
L=
\sum_q q \,b^{\dagger}_q b^{}_q.
\label{ChiralMomentum}
\end{equation}
In this definition $q$ is always positive, so that a state with linear quasimomentum $-q$ has chiral momentum $M-q$. The operator $L$ is therefore not time reversal.
In real space this operator has the form:
\begin{equation}
L=
\frac{(M-1)N}{2}-\frac{i}{2}\sum_{i\neq j} \frac{e^{-i\theta_{ij}}}{\sin\theta_{ij}} a^{\dagger}_i a^{}_j,
\label{ChiralMomentumRealSpace}
\end{equation}
with $N$ being the number operator.
It involves long-range tunneling with complex amplitude decaying in absolute value as the inverse of the chord distance $d_{ij}$.

Let me characterize the subspace 
$\mathcal{L}_k$
of many-body states of $k$-hard-core bosons with well defined $L$ and $N$. On the one hand, a state with well defined $L$ must be a symmetric homogeneous polynomial of degree $L$ in the variables $z_1,\dotsm, z_N$, such that the power of each coordinate does not exceed $M-1$, the maximum allowed chiral momentum for one particle. On the other hand, in order to describe $k$-hard-core bosons, this polynomial must vanish when $k+1$ particles coincide at the same position. For $k=1$, it follows that $(z_i-z_j)^2$ must be a factor for any pair of particles $i,j$, and $\Phi[z]=\prod_{i<j}(z_i-z_j)^2$ is therefore a factor. For $k=2$ the polynomial must vanish when three particles come together and therefore for any trio of particles $i,\ell, j$, one of the three terms $(z_i-z_j)^2$, $(z_i-z_\ell)^2$, or $(z_\ell-z_j)^2$ must be a factor. This implies that 
$\sum_{i,j\neq k,\ell}\prod_{i<j,k<\ell}(z_i-z_j)^2 (z_k-z_{\ell})^2$ is a factor of the wave function. For $N$ even, this factor can be written as $\mathcal{S} \{\Phi[z^{(1)}]  \Phi[z^{(2)}] \}$, the symmetrization over all possible ways of distributing the particles in two copies of the state $\Phi[z]$. Generalizing this result to the case of $k$-hard core bosons with $N=kn$, we obtain that the states in $\mathcal{L}_k$ are polynomials that have  $\widetilde\Psi^{(k)}=\mathcal{S} \{\Phi[z^{(1)}] \dotsm  \Phi[z^{(k)}] \}$ as a factor.

At filling factor $\nu=k/2$ there is only one state in $\mathcal{L}_k$ that is real. This state is $\Psi^{(k)}$ in Eq.\,(\ref{ClusterStates}), as can be proved in the following way. 
On the one hand, if $\Psi$ is real and has well defined $L$, it can be written as $\Psi=R \,\widetilde\Psi^{(k)}$, with $R$ a polynomial such that $R/R^*=\widetilde\Psi^{(k)*}/\widetilde\Psi^{(k)}$. Since $(z_i-z_j)^2=z_i^2z_j^2(z_i^*-z_j^*)^2$, we have that $R^*/R=\prod_i z_i^{2(N/k-1)}$, which for $\nu=k/2$
implies $R/R^*=\prod_i z_i^2$. Therefore $\prod_i z_i$ is a factor of $R$ and thus $\Psi^{(k)}$ is a factor of $\Psi$.  On the other hand, for any real wave function with well defined $L$ we have that
$L=(L+L^*)/2=MN/2-Mb^\dagger_0b^{}_0/2$. Since the factor $\prod_i z_i$ ensures that the power of each particle coordinate is larger than zero, we have that the number of particles in the $q=0$ momentum state vanishes and  $L=MN/2=N^2/2\nu$. Given that $\Psi^{(k)}$ is a polynomial of degree $N^2/k$, it follows that at $\nu=k/2$, $\Psi=\Psi^{(k)}$.

{\em Parent Hamiltonian for $k$-hard-core bosons with well defined total chiral momentum}.
Let me consider the following Hamiltonian:
\begin{equation}
H^{(k)}=P_k \left(L-\frac{MN}{2}\right)^2 P_k+ \text{c.c.}\,\,,
\label{ChiralMomHam}
\end{equation}
where $P_k$ is the projector onto the subspace of $k$-hard-core bosons and c.c denotes complex conjugation. By construction, the Hamiltonian (\ref{ChiralMomHam}) is positive definite, it is time reversal symmetric and has $\Psi^{(k)}$ as a ground state with zero eigenvalue.  For $k$-hard-core bosons at $\nu=k/2$ this ground state is non-degenerate, since for any $\Psi$ eigenstate of (\ref{ChiralMomHam}) with zero eigenvalue, we have that $LP_k\Psi=L^*P_k\Psi=MN/2\Psi$, properties that uniquely identify $\Psi^{(k)}$ within the subspace of $k$-hard-core bosons.

The Hamiltonian (\ref{ChiralMomHam}) can be written as $H^{(k)}=A_k^\dagger A^{}_k + B_k^\dagger B^{}_k + \text{c.c.}$, where $A_k=P_k(L-MN/2) P_k$, and $B_k=Q_kLP_k$,
with $Q_k=1-P_k$ being the projector onto the orthogonal subspace. The Hamiltonian $B_k^\dagger B^{}_k+\text{c.c.}=(k+1)H_B^{(k)}$, with $H_B^{(k)}$ the Hamiltonian in Eq.\,(\ref{Ham3Body}). It involves virtual processes in which a particle leaves the $k$-hard-core bosonic subspace and returns. The Hamiltonian $A_k^\dagger A^{}_k+\text{c.c.}$ can be written as 
$(C_k^\dagger C^{}_k)^2+D_k^\dagger D^{}_k$, where $C_k=\sqrt{M}P_kb_0P_k=\sum_i a^{(k)}_{i}$ is the projection onto the subspace of $k$-hard-core bosons of the annihilator operator in the state of zero momentum. The operator
$D_k=P_k (L-L^*) P_k=-i\sum_{i\neq j}\cot \theta_{ij} a^{\dagger(k)}_{i} a^{(k)}_{j}$, involves tunneling processes of complex amplitude within the subspace of $k$-hard core bosons. Altogether, the Hamiltonian (\ref{ChiralMomHam}) can be written as:
\begin{equation}
H^{(k)}= \! \! \!\sum_{i \neq j, \ell \neq m} \!\!\!\!
V_{i m}^{\ell j}\,\,a^{\dagger(k)}_{i} a^{\dagger(k)}_{\ell} a^{(k)}_{m}  a^{(k)}_{j} +
\sum_{i \neq j} I_{ij}\,
a^{\dagger(k)}_{i} a^{(k)}_{j},
\label{Ham4Body}
\end{equation}
where $V_{im}^{\ell j}=1-\cot \theta_{i m} \cot \theta_{\ell j}\sim 1/d_{im}d_{\ell j}$, and $I_{ij}=\sum_\ell V_{i\ell}^{\ell j}=\sin^{-2}\theta_{ij}+2(M-1) \sim d^{-2}_{ij}$ \cite{Note}. 

This parent Hamiltonian can be simplified to a Hamiltonian containing only interactions involving three lattice sites. Such a simplified Hamiltonian is the one in Eq.\,(\ref{Ham3Body}), with couplings $J_{i\ell j}=V_{i \ell}^{\ell j}$.
The state $\Psi^{(k)}$ is an eigenstate of this Hamiltonian with eigenvalue zero, since we have
$B_k\Psi^{(k)}=B_k^*\Psi^{(k)}=Q_kLP_k\Psi^{(k)}=Q_kL^*P_k\Psi^{(k)}=MN/2Q_k\Psi^{(k)}=0$. Moreover, the state $\Psi^{(k)}$ is the only ground state satisfying $C_k\Psi=0$, as can be seen as follows.
Let us consider a ground state $\Psi$ of the Hamiltonian in Eq.\,(\ref{Ham3Body}). Without loss of generality we can assume that this state is real. If $C_k\Psi=0$ we have that $(L+L^*)\Psi=(P+Q)(L+L^*)\Psi=P(L+L^*)P\Psi=(MN-C^{\dagger}_kC^{}_k)\Psi=MN\Psi$, and thus $\langle\Psi\vert L\vert\Psi\rangle=MN/2$. Since the maximum eigenvalue of $L$ within the $k$-hard-core bosonic subspace is $MN/2$, it follows that $\Psi$ is an eigenstate of $L$, and that therefore it must coincide with $\Psi^{(k)}$.
The condition $C^{}_k \Psi=0$ can be better understood by turning to a language of spins, in which it is revealed as the condition of being a total singlet. By imposing invariance under global rotations the corresponding spin state can be selected as a unique ground state.

{\em k/2-spin liquids and their parent Hamiltonians}. 
The Hilbert space of $k$-hard-core bosons is isomorphous to the one of $\frac{k}{2}$-spins. Let me define the following mapping between bosonic and spin operators:
\begin{align}\label{OperatorCorrespondence}
\nonumber a^{(k)}_i & \longrightarrow  S^-_i \\ 
a^{\dagger(k)}_i & \longrightarrow  \widetilde{S}^+_i=T_iS^+_i \\
\nonumber a^{\dagger(k)}_i a^{(k)}_i & \longrightarrow S^z_i+\frac{k}{2},
\end{align}
where $T_i=(1+\frac{k}{2}-S_i^z)^{-1}$, and $S_i^+,S_i^-,S_i^z$ are $\frac{k}{2}$-spin operators of the $i$th spin.
This mapping is not unitary (except for $k=1$), but preserves the commutation relations of the $k$-hard-core bosonic operators, so that $[a^{(k)},a^{\dagger(k)}]=[S^-,\widetilde{S}^+]$. Therefore, given an arbitrary state of $k$-hard core bosons $\ket{\Psi}=F[a^{\dagger(k)}]\ket{0}$, and $G(a^{\dagger(k)},a^{(k)})$ an annihilator of this state, it follows that the mapping in Eq.\,(\ref{OperatorCorrespondence}) will convert it into a spin state $\ket{\Psi_S}=F[\widetilde{S}^{+}]\ket{-}$ that is annihilated by the operator $G(\widetilde{S}^{+},S^{-})$.

In particular, the transformation in Eq.\,(\ref{OperatorCorrespondence}) maps the bosonic state $\vert\Psi^{(k)}\rangle=\sum_{x_1,\dotsc, x_{kn}} \!
\Psi^{(k)}[z_1,\dotsc,z_{kn}] \,\,\,
a^{\dagger}_{x_1} \dotsm a^{\dagger}_{x_{kn}} \ket{0}$ onto the spin state in Eq.\,(\ref{SpinState}). Since $\Psi^{(k)}$ is annihilated both by $\sum_i a^{\dagger(k)}_ia^{(k)}_i-k/2$ and $C_k=\sum_ia^{(k)}_i$, it follows that the corresponding spin state is annihilated by $\sum_iS^z_i$ and $\sum_iS^-_i$, and is therefore a total singlet. 
Moreover, using the fact that $\Psi^{(k)}$ is the only ground state of Hamiltonian (\ref{Ham3Body}) with $\nu=k/2$ that is annihilated by $C_k$, we can conclude (see supplementary material) that $\Psi_{\text{spin}=k/2}$ is the only total singlet ground state of the following spin Hamiltonian:
\begin{equation}
H_{\text{spin}=k/2}=\sum_{ \ell \neq i, \ell \neq j}
\frac{\cos \theta_{ij}}{\sin \theta_{i \ell} \sin \theta_{j \ell}} 
S^{+}_{i} P^{(k/2)}_{\ell} S^{-}_{j}.
\label{Ham3Spin}
\end{equation}
Here $P^{(k/2)}_{\ell}$ projects the state of the $\ell$th spin onto the state with maximum $z$ component $S^z_\ell=k/2$. This Hamiltonian is positive definite. Its rotationally invariant form is therefore also positive definite and has $\Psi_{\text{spin}=k/2}$ as a unique ground state.

{\em Spin-1/2 Haldane-Shastry Hamiltonian}. For $k=1$ the Hamiltonian (\ref{Ham3Spin}) is the spin-$1/2$ Hamiltonian 
$\sum_{ \ell \neq i \neq j} J_{i\ell j}[S_i^{x}S_j^{x}+S_i^{y}S_j^{y}](S_\ell^z+\frac{1}{2})+\sum_{i\neq j}K_{ij}(S_i^z+\frac{1}{2})(S_j^z+\frac{1}{2})$, with $J_{i\ell j}$, $K_{ij}$ defined above.
By imposing invariance under global rotations, the terms involving three sites, which have odd parity, cancel, and the Hamiltonian is converted (up to an irrelevant energy shift) into the Haldane-Shastry Hamiltonian:
\begin{equation}
H_{\text{spin}=1/2}=\sum_{i<j}
\frac{1}{\sin^2 \theta_{ij}}\vec{S}_i \cdot \vec{S_j}.
\label{HaldaneShastry}
\end{equation}
{\em Spin-1 parent Hamiltonian}. For $k=2$ the Hamiltonian (\ref{Ham3Spin}), after imposing invariance under parity, is converted into the spin-$1$ Hamiltonian:
\begin{align}
\nonumber H_{\text{spin}=1}=& \sum_{ i \neq \ell \neq j,i<j} J_{i\ell j}\,[S_i^{x}S_j^{x}+S_i^{y}S_j^{y}][S_\ell^z]^2 \,+ \\ &\sum_{i< j}K_{ij}([S_i^z+S_j^z]^2-S_i^zS_j^z[1+S_i^zS_j^z]).
\label{Ham3Spin1}
\end{align}
The second term in Eq.\,(\ref{Ham3Spin1}) can be rewritten as $-[P_i^{+1}P_j^{-1}+P_i^{-1}P_j^{+1}]$, with $P_i^{+1(-1)}$ projecting the state of the $i$-th spin onto the state $\ket{+1}(\ket{-1})$. This reveals effective long range attractive interaction between states $+1$ and $-1$. By imposing invariance under global rotations, the Hamiltonian in Eq\,(\ref{Ham3Spin1}) is transformed into:
\begin{align}
H_{\text{spin}=1}^{}=H_{\text{spin}=1}^{(3)} + 
\sum_{i < j}K_{ij}[\alpha \vec{S}_i \cdot \vec{S}_j + \beta [\vec{S}_i \cdot \vec{S}_j]^2],
\label{Ham3Spin1Inv}
\end{align}
with $H_{\text{spin}=1}^{(3)}$ the Hamiltonian in Eq.\,(\ref{HamSpin1}), and $\alpha, \beta$ rational numbers. 

{\em Non-Abelian spin liquids as clusters of RVB states}. 
To get insight into the spin liquid ground states we have introduced, it is helpful to notice that their corresponding bosonic states in Eq.\,(\ref{ClusterStates}) can be written as:
\begin{align}
	\vert \Psi^{(k)}\rangle=\left[f(a^{\dagger})\right]^k\ket{0},
\label{ProductForm}
\end{align}
where $f(a^{\dagger})=\sum_{x_1,\dotsc, x_{n}} 
\Psi^{(1)}[z_1,\dotsc,z_{n}]\,a^{\dagger}_{x_1} \dotsm a^{\dagger}_{x_{n}} $ characterizes the Laughlin bosonic state.
In particular, for $k=1$, $\ket{\Psi_{\text{spin}=1/2}}=f(S_{\text{s}=\frac{1}{2}}^{+})\ket{-}_{\frac{1}{2}}$ is the Haldane RVB state. For $k=2$, the Pfaffian spin-$1$ liquid is
$\ket{\Psi_{\text{spin}=1}}=[f(\widetilde{S}^{+}_{\text{s}=1})]^2\ket{-}_{\text{s}=1}$. Introducing two effective $1/2$ spins per lattice site, we have that for each lattice site 
$[\widetilde{S}_{\text{s}=1}^{+}]^m\ket{-1}= \mathcal{P}^{(\text{s}=1)}[S^+_{\text{s}=\frac{1}{2}}]^{m'}\ket{-1/2}
\otimes[S^+_{\text{s}=\frac{1}{2}}]^{m-m'}\ket{-1/2}$, where $\mathcal{P}^{(\text{s}=1)}$ is the projector onto the symmetric subspace (the space of triplets, with $s=1$) of these two effective spins, and $m,m'=0,1,2$, with $m-m'>0$. Thus we can write:

\begin{align}
	\ket{\Psi_{\text{spin}=1}}=\prod_i \mathcal{P}_i^{(\text{s}=1)} f(S^+_{\text{s}=\frac{1}{2}})\ket{-}_{\frac{1}{2}} \otimes f(S^+_{\text{s}=\frac{1}{2}})\ket{-}_{\frac{1}{2}},
\label{Spin1RVB}
\end{align}
revealing its hidden organization in two indistinguishable RVB states.

In conclusion, I have presented a family of one dimensional bosonic and spin liquids described by non-Abelian fractional quantum Hall many-body wave functions. These states exhibit hidden global order, associated to the organization of the particles or spins in a collection of identical indistinguishable clusters with well defined total chiral momentum. For the case of bosonic particles, each cluster is a Laughlin state. For the case of spins, each cluster is a Haldane RVB state. 
This work demonstrates the stability of these states by constructing a family of exact parent Hamiltonians involving long-range interactions between a few number of particles or spins. 
These Hamiltonians could be realized in experiments with polar molecules stored in optical lattices \cite{Zoller}, where the $k$-hard-core regime could be achieved through strong on-site repulsion.
The non-Abelian character of these states is revealed in the presence of non-Abelian excitations. 
The analysis of the gap and the excitations of the Hamiltonians presented here is the subject of an upcoming work in which it will be shown that the hidden symmetry (particle-hole symmetry, in the case of bosons, and parity, in the case of spins) leads to the existence of a finite energy gap. 
The realization of these phases would open a door to the observation of non-Abelian phases and anyons in one-dimensional systems.

In addition, the states and Hamiltonians presented here can provide us with a useful description at an exact level of one-dimensional Haldane insulators \cite{Altman2006,Altman2008} and Haldane spin gapped phases exhibiting string order. 
Moreover, the class of many-body wave functions presented here, puts forward a broader class of functions, consisting of indistinguishable arbitrary {\em two-dimensional} RVB states, which are promising as trial wave functions to describe non-Abelian spin liquids in two dimensions. Finally, the approach followed here to derive the parent Hamiltonians by imposing conservation of the global quantum numbers associated to the hidden symmetries, sketches a path for the investigation of the intriguing relation between the microscopic interactions in a quantum many-body system and the emergent non-Abelian topological order.

After submitting this manuscript, it was brought to my attention that closely related Hamiltonians and wave functions have been recently derived in \cite{Sierra2011, Greiter2011}.


\begin{thebibliography}{10}
\bibitem{Note}
Along the whole manuscript, energy couplings are given in units of $4/M^2$, with $M$ being the number of lattice sites.
\end{thebibliography}
\end{document}